\documentclass[twocolumn,twoside,slac]{revtex4}
\usepackage{graphicx}
\usepackage{fancyhdr}
\begin{document}

\title{SCRAM: Software configuration and management for the LHC Computing Grid project.}

%

\author{J.P. Wellisch}
\affiliation{CERN}
\author{C. Williams}
\affiliation{CERN}
\author{S. Ashby}
\affiliation{CERN}

\begin{abstract}
Recently SCRAM (Software Configuration And Management) has been adopted by the
applications area of the LHC computing grid project as baseline configuration
management and build support infrastructure tool.

SCRAM is a software engineering tool, that supports the configuration management
and management processes for software development. It resolves the issues of
configuration definition, assembly break-down, build, project
organization, run-time environment, installation, distribution, deployment,
and source code distribution. It was designed with a focus on supporting
a distributed, multi-project development work-model.

We will describe the underlying technology, and the solutions SCRAM offers to
the above software engineering processes, while taking a users view of the
system under configuration management. 
\end{abstract}

\maketitle

\thispagestyle{fancy}

\section{Configuration management concepts and their realization in SCRAM}
Configuration management with SCRAM is based on the following concepts:
\begin{itemize}
\item PBS
\item Product specification
\item Product versioning
\item Configuration definition and versioning
\item Product-wise configuration selection
\item Assembly break-down.
\end{itemize}
Here the PBS is the set of products, tools, etc. that are used in all software systems. These may be compilers like gcc, external packages like zlib or root, or experiment owned software like the reconstruction code. 

SCRAM product specifications are expected to be collected in a repository of their own, that enables versioning, and distributed access. There is exactly one specification file per product, which gives for each version of the product a complete description the product and the environment, compile time and run time, it needs for being used. This can relate to include paths, library paths, runtime settings, locations for graphics resources, license file-names, etc..

Product versioning is not done by SCRAM. It relies in this respect on a standard versioning system, like cvs, or simply the version numbers of external components, as delivered.

The configuration is then defined as an inclusive and consistent set of product specifications and versions, that are referred to in a single file, the configuration file. This configuration file in turn is versioned using a standard versioning system, like cvs. It can be referred to by name and version from any SCRAM managed software development project in the project's 'Requirements' file. To realize the assembly break down, each SCRAM managed project then selects from the configuration, by name, the tools/products it depends on. The version numbers of these tools will then be taken as specified in the selected version of the configuration file. The requirements document itself is parts of a project's release. 

\section{Functionality and design}
The principal goal of SCRAM is to provide developers located in different institutions and using differing operational environments with
an easy way to use consistent development environments. By doing this, SCRAM  enables massively parallel and distributed software development in a multi-project environment. It does this by applying strict configuration management. In addition, it provides a build system, and a mechanism for source code distribution.

SCRAM is mapping the application's requirements onto a given target operational environment. It extracts information on the target system, and maps the configuration requirements of each product used onto actually installed components in the local operational environment.

The SCRAM managed project has three areas of concern. Source code distribution which is governed by the BootStrapFiles, configuration management governed by the Requirements documents, and the builds, governed by the BuildFiles.

SCRAM itself is written in object oriented perl, for easy installation.
\subsection{SCRAM documents and parsing}
All SCRAM documents are written in a XML style language, and SCRAM contains the ActiveDoc parser, to handle these documents. 

Each document is typed and versioned in turn, like in 
{\small
\begin{verbatim}
<doc type=a:b version=2.1>
\end{verbatim}
}
Functions may then map any document tag onto any object method. Multi-parsing is enabled, hence documents can refer to other documents. The syntax is based on a URL retrieval mechanism. Currently {\rm cvs:} and {\rm file:} types are supported, and an interface for plugging in new types is provided. Documents can be inlined using  
{\small
\begin{verbatim}
<inline url="MyDocumentToInline">.
\end{verbatim}
}
A cache mechanism for already fetched documents optimizes access and retrieval times.

The document classes available are ActiveDoc, SimpleURLDoc, and SimpleDoc. Here SimpleDoc is the base-class, has the parsing kernel, enables multi-parsing, and grouping of tags. SimpleURLDoc adds URL retrieval mechanisms, document versioning, as well as the URL cache. ActiveDoc finally adds preprocessing and type activation including the object store.

\subsection{Client project installation}
The mechanism for installation of a SCRAM managed product is described in its BootStrapFile. This file is part of any release of a SCRAM managed project. A client will start the installation procedure by downloading the projects BootStrapFile through WWW. SCRAM then assembles the components described in the BootStrap document into a central installation area. In case a Requirements document is specified in the released project, SCRAM attempts to find the selected products in the client's  operational environment. If provided, SCRAM uses the specification in a site description file to achieve this. For products that cannot be automatically mapped completely, the installer will be asked to provide the information necessary to conclude the resource mapping on the command line.
Depending on the nature of the project, the installer then may have to issue a build command, {\bf scram build}. Once the build was successful and the installation was verified, issuing the {\bf scram install} command will make the central installation available for dependent work by developers and users. It is now ready for use.

\subsection{Requirement and product specification documents}
The product specification documents serve the purpose to make it such that users and developers do not need to know about the environment needed by the application and development areas. One of the more complicated real-life examples of a tool description file is given below:
{\small
\begin{verbatim}
<doc type=BuildSystem::ToolDoc version=1.0>
<Tool name=Boost version=1.28.0>
<info url=http://www.boost.org></info>
<Lib name=boost_thread>
<Client>
<Environment name=BOOST_BASE>
  The top of the Boost distribution.
</Environment>
<Environment name=LIBDIR type=lib></Environment>
<Environment name=INCLUDE></Environment>
</Client>
<External ref=sockets version=1.0>
We need the sockets libs
</External>
<Environment name=LD_LIBRARY_PATH value=$LIBDIR 
             type=Runtime_path></Environment>
</Tool>
<Tool name=Boost version=1.29.0>
<info url=http://www.boost.org></info>
<Lib name=boost_thread>
<Client>
<Environment name=BOOST_BASE>
The top of the Boost distribution.
</Environment>
<Environment name=LIBDIR type=lib></Environment>
<Environment name=INCLUDE></Environment>
</Client>
<External ref=sockets version=1.0>
We need the sockets libs
</External>
<Environment name=LD_LIBRARY_PATH value=$LIBDIR 
             type=Runtime_path></Environment>
</Tool>
<Tool name=Boost version=1.30.0>
<info url=http://www.boost.org></info>
<Lib name=boost_thread>
<Client>
<Environment name=BOOST_BASE>
The top of the Boost distribution.
</Environment>
<Environment name=LIBDIR type=lib></Environment>
<Environment name=INCLUDE></Environment>
</Client>
<External ref=sockets version=1.0>
We need the sockets libs
</External>
<Environment name=LD_LIBRARY_PATH value=$LIBDIR 
             type=Runtime_path></Environment>
</Tool>
\end{verbatim}
}

This example specifies the product boost for a set of versions of this product.
It specifies for each version the name of the product, where to find product information, and the names of the product's libraries. It enables search for library and include paths, and specifies the external dependencies of the product, in this case the sockets library.

One product specification file is maintained per product, ideally by the author or an expert of this product. Users and developers in turn have no need to know any of the details of the product's implementation and installation to be able to use it.

The product specification files will be collected into a versioned configuration file, using {\bf require} tags. Below a snippet of such a configuration file:
{\small
\begin{verbatim}
...
<Architecture name=SunOS__5>
  <require name=f77  version=4.2		url="cvs:?module=SCRAMToolBox/Fortran/SunF77">
  </require>
  <require name=CC   version=5.4		url="cvs:?module=SCRAMToolBox//CXX/SunCC">
  </require>
</Architecture>
<Architecture name=Linux__2.4>
  <require name=gcc3 version=3.2		url="cvs:?module=SCRAMToolBox/CXX/gcc3">
  </require>
  <require name=gcc  version=2.95.2	url="cvs:?module=SCRAMToolBox/CXX/gcc">
  </require>
  <require name=g77  version=0.5.24	url="cvs:?module=SCRAMToolBox/Fortran/g77">
  </require>
  <require name=icc  version=7.0	        url="cvs:?module=SCRAMToolBox/CXX/icc">
  </require>
</Architecture>
  <require name=LHCxx  version=5.0.3	url="cvs:?module=SCRAMToolBox/LHCxx/LHCxx">
  </require>
  <require name=Qt     version=3.1.2	url="cvs:?module=SCRAMToolBox/LHCxx/Qt">
  </require>
  <require name=CLHEP  version=1.8.0.0	url="cvs:?module=SCRAMToolBox/LHCxx/CLHEP">
  </require>
...
\end{verbatim}
}

It shows how the various products with concrete versions are grouped together
into one configuration. Dependencies on the operational environment, like in the case of
compilers, are taken into account
through {\rm \bf Architecture} tags.

Any SCRAM managed project will have a Requirements document, 
that specifies the configuration and configuration version it wishes to use.
In this document, the project will then select by name the products used.
An example snippet below:
{\small
\begin{verbatim}
<doc type=BuildSystem::Requirements version=2.0>
<base url="cvs://cmscvs.cern.ch/.../SCRAMToolBox
       ?auth=...&user=...&version=CMS_68_2">
<include url="cvs:?module=.../CMSconfiguration">
<Architecture name=SunOS__5.8>
<select name=CC>
<select name=f77>
</Architecture>
<Architecture name=Linux__2.4>
<select name=gcc3>
<select name=g77gcc3>
</Architecture>
<select name=COBRA>
<select name=IGUANA>
<select name=CMSToolBox>
<select name=Geometry>
<select name=AIDA>
<select name=AIDA_Dev>
<select name=AIDA_XMLStore>
<select name=AIDA_AnalysisFactory_native>
<select name=AIDA_Tree_native>
\end{verbatim}
}
Again {\bf Architecture} tags are used to resolve dependencies on the operational environment.
Since the version selection is done centrally in the configuration, a multi-project
environment is kept consistent in the most natural manner.

\subsection{The distributed development model}
SCRAM assumes a work-model, in which development is done in multiple projects, and for each
project on multiple, internationally distributed sites. It is assuming a rather rapid release cycle,
and that an installation of the project versions to be further developed is centrally
available at each site. Developers will have a local development area, drawing resources from
this central installation, while only the part that a developer wishes to change
is kept locally. It is the central installation that maps the projects requirements onto the local system's resources, contains a 
complete set of source code and build products, provides for the definition of the environment
associated to the project, has the URL and object cache, but is still self-contained, and
hence movable without breaking. It also defines the context of all functions of SCRAM.

The developer areas are linked to the central installation, and use all elements of this
installation that were not changed by the developer, resulting in a significant saving of
both time and computing resources. 

Easy look-up of available central installations is
provided, and the context of the development area is defined by the associated installation.
Changing the development area, using {\bf cd}, will result in a change of the environment.
This enables easy and transparent use of multiple releases and projects by an
individual developer. 

At the time of writing, the development area is tightly controlled by the build system.

\subsection{The build system}
At the time of writing, the build system is undergoing major changes. A description of these
is deferred to a later publication.

\section{The users view}
\subsection{Development areas and central installations}
The user of SCRAM is provided with an isolated and well defined environment so he can focus
on development of code, while the environment is defined by the central installation he bases his work on. 
The user will find the available installations by using the {\bf scram list} command 
as in the following:
{\small
\begin{verbatim}
scram list ORCA

Listing installed projects....

------------------------------------
| Project  | Version  |  Location  |
------------------------------------
...
ORCA 7_1_2 --> /afs/cern.ch/.../ORCA_7_1_2
ORCA 7_1_3 --> /afs/cern.ch/.../ORCA_7_1_3
...
Projects available for platform >> Linux__2.4 <<
\end{verbatim}
}

He will then want to create the desired developers area using the {\bf scram project} command,
like in:
{\small
\begin{verbatim}
scram project ORCA 7_1_3
\end{verbatim}
}
This will create a development area that refers to the central installation of a project
called ORCA with version 7\_1\_3. 

The structure of the development area is such that it contains a {\bf src} area to be used by the user for changing source code, a {\bf config} area that
contains the information on the configuration, {\bf lib} and {\bf bin} areas for build
products, a {\bf logs} area for log file storage, a {\bf tmp} area that is the working area of the
build system, and a {\bf .SCRAM} area, which is used by SCRAM for
administration files. The real working area hence is the {\bf src} directory.

\subsection{Product description management}
As described earlier, all product descriptions live in a repository. If a developer wishes
to find out what products he depends on, he can issue the {\bf scram tool list}
command to query the installation. This will yield the complete list of products used, with their version in the
configuration, and the version used by the development area. Normally, the two will be
identical.
{\small
\begin{verbatim}
scram tool list

Tool list for location .../ORCA_7_1_2
++++++++++++++++++++++++++++++++++++++++++++++++++
...
 gcc                  2.95.2     (default=2.95.2)
 g77                  0.5.24     (default=0.5.24)
 qt                   3.0.5      (default=3.0.5)
 clhep                1.8.0.0    (default=1.8.0.0)
 heputilities         0.7.1.1    (default=0.7.1.1)
 gemini               1.3.0.3    (default=1.3.0.3)
 openinventor         3.1.1      (default=3.1.1)
...
\end{verbatim}
}
If a user wishes to use a product not in the configuration, or wishes to change a products
version, he can issue the {\bf scram setup} command, like in
{\small
\begin{verbatim}
scram setup htl 1.5 cvs://...&module=HTL
\end{verbatim}
}
SCRAM will in turn try to find the specified product/version combination on the developer's system, may ask questions on the command line to obtain additional information, may provide hints for
installation of the product, etc.. 

To find out details on the environment a particular product has, the {\bf scram tool info}
command can be used. It will simply retrieve and display 
the information specified in the product specification
for the version of the product in current use.

Developing product specifications is done by editing the specification files, and running
{\bf scram setup} to initiate the changes for the local development area. Normally this
should be done in the context of configuration management.
To make local changes useful outside the individual developer's area, they need to be made
available such that the URL access mechanism can be used with one of the supported tags.

\subsection{The environment}
The runtime environment is set up by entering the directory structure of the development area, and
issuing the command {\bf eval `scram runtime -csh`}. This will set the environment for the
current development area consistent with the requirements of the corresponding 
central installation. It also will roll back any environment settings that may stem from
different development areas this user works on, so switching areas is easy.

In case different applications built in the same project context require different runtime
settings, these can be described in an application environment file. SCRAM will read this
file, and adjust the environment accordingly.

\section{Part of a component system}
SCRAM was conceived to be part of a component system for process support, and hence defines interfaces to other parts of a general process support infrastructure. The processes and process elements considered are code management, change management, binary distribution, dependency analysis and QA, software integration, and validation and verification.

\section{Underlying technology}
With the decision of building SCRAM as a scripted instrument, the requirements for external products have been kept to a strict minimum. The exhaustive list of these products is {\bf perl}, a {\bf make} tool, and a {\bf shell}.

\begin{acknowledgments}
The authors wish to thank CERN for their support.
\end{acknowledgments}

\end{document}